\documentclass[preprint,amsmath,showpacs]{revtex4}
\usepackage{graphicx}% Include figure files
\usepackage{dcolumn}% Align table columns on decimal point
\usepackage{bm}% bold math

\def\bee{\begin{eqnarray}}
\def\eee{\end{eqnarray}}

\begin{document}
\title{Determinations of flavor ratios and flavor transitions of astrophysical
neutrinos\footnote{Presented at 35th International Conference of High Energy Physics - ICHEP2010,\\
        July 22-28, 2010\\
        Paris France}}

\author{Guey-Lin Lin$^{1,2}$}
\author{T. C. Liu$^{2}$}
\author{Kwang-Chang Lai$^{2,3}$}
\author{M. A. Huang$^{2,4}$}
\affiliation
        {$^{1}$Institute of Physics, National Chiao-Tung University, Hsinchu 300, Taiwan}
\affiliation{$^{2}$Leung Center for Cosmology and Particle
Astrophysics, National Taiwan University, Taipei 106, Taiwan}
\affiliation{$^{3}$Physics Group, Center for General Education,
Chang Gung University, Kwei-Shan 333, Taiwan}
\affiliation
        {$^{4}$Department of Energy and Resources, National United University, Miao-Li 36003, Taiwan}
\date{\today}

\begin{abstract}
We argue that an effective flavor discrimination in neutrino
telescopes is the key to probe the flavor ratios of astrophysical
neutrinos at the source and flavor transition mechanisms of these
neutrinos during their propagations from the source to the Earth. We
first discuss how well one can reconstruct the flavor ratios of
astrophysical neutrinos at the source. We then discuss how to probe
flavor transition mechanisms of propagating astrophysical neutrinos.
In this regard, we propose a model independent parametrization for
neutrino flavor transitions, with the neutrino oscillation as a
special case.
\end{abstract}

\maketitle

\section{Determining Flavor Ratios of Astrophysical Neutrinos}
It has been demonstrated that \cite{Beacom:2003nh} the event ratio
of muon tracks to showers can be measured to a $10\%$ accuracy for a
decade of data taking in IceCube with a neutrino flux in the order
of Waxman-Bahcall bound \cite{Waxman:1998yy}. The above accuracy in
measuring track to shower ratio can be translated into a comparable
accuracy for determining the flux ratio $R\equiv\phi
(\nu_{\mu})/\left(\phi (\nu_{e})+\phi (\nu_{\tau})\right)$. We have
pointed out that a well determined $R$ is still not sufficient for
constraining the neutrino flavor ratio at the source, which is
characterized by the vector
$\Phi_0=(\phi_{0}(\nu_e),~\phi_{0}(\nu_{\mu}),~\phi_{0}(\nu_{\tau}))^T$
with \cite{comment}
\begin{eqnarray}
&&\phi_{0}(\nu_e)+\phi_{0}(\nu_{\mu})+\phi_{0}(\nu_{\tau})=1, \nonumber \\
&&\phi_{0}(\nu_{\alpha})\geq0, \hspace{.5mm}{\rm for}\hspace{1mm}
\alpha=e,\mu,\tau,\label{norma}
\end{eqnarray}
where each $\phi_{0}(\nu_{\alpha})$ is the sum of neutrino and
antineutrino fluxes. On the other hand, we have shown that
\cite{Lai:2009ke} it is possible to discriminate between an
astrophysical pion source with $\Phi_0=(1/3,1/2,0)^T$ and a muon
damped source with $\Phi_0=(0,1,0)^T$ \cite{Pakvasa:2004hu} at the
$3\sigma$ level, provided that the separation between the tau shower
and the electron shower, namely the measurement of the parameter
$S\equiv\phi (\nu_{e})/\phi(\nu_{\tau})$, can be done as effectively
as $R$ in neutrino telescopes. To reach this conclusion, we have
taken into account the following uncertainties ($1\sigma$ ranges) of
neutrino mixing parameters \cite{GonzalezGarcia:2007ib}
\begin{eqnarray}
\sin^2\theta_{12}=0.32^{+0.02}_{-0.02}, \,
\sin^2\theta_{23}=0.45^{+0.09}_{-0.06}, \, \sin^2\theta_{13}< 0.019.
\label{bestfit}
\end{eqnarray}
We conclude that tau neutrino identification is crucial for
developing the neutrino flavor astronomy.

We also consider the detection of very high energy astrophysical
neutrinos, i.e., $E_{\nu}> 33$ PeV. In such an energy range, the tau
lepton originated from the tau neutrino behaves more like a track
rather than a shower while the electron neutrino only give rises to
a shower signature. Therefore the more appropriate flux ratio
parameters in such a case are $R^{\prime} \equiv\phi
(\nu_{e})/\left(\phi (\nu_{\mu})+\phi (\nu_{\tau})\right)$ and
$S^{\prime}\equiv\phi (\nu_{\mu})/\phi(\nu_{\tau})$. We found that
$R^{\prime}$ is a more sensitive parameter than $R$ for
reconstructing the neutrino flavor ratio at the source, if both
parameters are measured to the same accuracy \cite{lhl_2010}.
Furthermore, it is interesting to note that a further measurement on
$S^{\prime}$ does not improve the reconstruction of initial neutrino
flavor ratios. In fact, due to the approximate
$\nu_{\mu}-\nu_{\tau}$ symmetry
\cite{Balantekin:1999dx,Harrison:2002et}, $S^{\prime}$ is always
close to $1$ irrespective of the initial neutrino flavor ratio. We
note that $R^{\prime}$ can be measured in the radio extension of
IceCube detector \cite{Allison:2009rz}, which aims for detecting
very high energy neutrinos such as those produced by the
interactions between ultrahigh energy cosmic rays and cosmic
microwave background radiations.

\section{Probing Flavor Transitions of Astrophysical Neutrinos}

The effect of neutrino flavor transition processes occurring between
the astrophysical source and the Earth is represented by the matrix
$P$ such that
\begin{equation}
\Phi=P\Phi_0, \label{P_representation}
\end{equation}
where $\Phi=(\phi (\nu_e),~\phi (\nu_{\mu}),~\phi (\nu_{\tau}))^T$
is the flux of neutrinos reaching to the Earth. We note that our
convention implies $P_{\alpha\beta}\equiv P(\nu_{\beta}\to
\nu_{\alpha})$. The matrix $P$ can be easily calculated for standard
neutrino oscillation model. However we should keep $P$ as general as
possible for accommodating other flavor transition models. It is
convenient to parametrize the initial flux of neutrinos by
\cite{Lai:2009ke}
\begin{equation}
\Phi_0=\frac{1}{3}{\mathbf V}_1+a{\mathbf V}_2+b{\mathbf V}_3,
\label{source_flux}
\end{equation}
where ${\mathbf V}_1=(1,1,1)^T$, ${\mathbf V}_2=(0,-1,1)^T$, and
${\mathbf V}_3=(2,-1,-1)^T$. The ranges for $a$ and $b$ are
$-1/3+b\leq a\leq 1/3-b$ and $-1/6\leq b\leq 1/3$. Likewise,one can
also write the flux on Earth in the same basis
\begin{equation}
\Phi=\kappa {\bf V}_1+\rho {\bf V}_2+\lambda {\bf V}_3.
\label{measured_flux}
\end{equation}
It is easy to show that
\begin{equation}
\left(
  \begin{array}{c}
     \kappa \\
    \rho \\
    \lambda \\
  \end{array}
\right) = \left(
  \begin{array}{ccc}
    Q_{11} & Q_{12} & Q_{13} \\
     Q_{21} & Q_{22} & Q_{23} \\
     Q_{31} & Q_{32} & Q_{33} \\
  \end{array}
\right) \left(
  \begin{array}{c}
     1/3 \\
    a \\
    b \\
  \end{array}
\right)
 , \label{new_basis}
\end{equation}
where $Q={\mathbf A}^{-1}P{\mathbf A}$ with
\begin{equation}
{\mathbf A}= \left(
  \begin{array}{ccc}
    1 & 0 & 2 \\
     1 & -1 & -1 \\
     1 & 1 & -1 \\
  \end{array}
\right). \label{evector}
\end{equation}
In other words, $Q$ is related to $P$ by a similarity transformation
where columns of the transformation matrix ${\mathbf A}$ correspond
to vectors ${\mathbf V}_1$, ${\mathbf V}_2$, and ${\mathbf V}_3$,
respectively.

For the flux-conserving case during the neutrino propagations, one
has $\kappa=1/3$ which requires $Q_{11}=1$ and $Q_{12}=Q_{13}=0$,
since the coefficients $a$ and $b$ are arbitrary. Furthermore, in
the exact $\nu_{\mu}-\nu_{\tau}$ symmetry limit, one can show that
$Q_{21}=Q_{22}=Q_{23}=Q_{32}=0$. Therefore, under these two
assumptions, there are only two free parameters, $Q_{31}$ and
$Q_{33}$, for classifying all possible neutrino flavor transition
models. This is a great advantage of $Q$ matrix parametrization for
describing neutrino flavor transitions. The determinations of
$Q_{31}$ and $Q_{33}$ by neutrino telescopes are discussed in
Ref.~\cite{Lai:2010tj}.


\begin{thebibliography}{99}

\bibitem{Beacom:2003nh}
J.~F.~Beacom, N.~F.~Bell, D.~Hooper, S.~Pakvasa and T.~J.~Weiler,
%``Measuring flavor ratios of high-energy astrophysical neutrinos,''
Phys.\ Rev.\  D {\bf 68}, 093005 (2003) [Erratum-ibid.\  D {\bf 72},
019901 (2005)].
%[arXiv:hep-ph/0307025].
%%CITATION = PHRVA,D68,093005;%%
%
\bibitem{Waxman:1998yy}
E.~Waxman and J.~N.~Bahcall,
%``High energy neutrinos from astrophysical sources: An upper bound,''
Phys.\ Rev.\  D {\bf 59}, 023002 (1998).
%[arXiv:hep-ph/9807282].
%%CITATION = PHRVA,D59,023002;%%
%
\bibitem{comment}
The subscript $0$ is used to denote the neutrino flux at the source.
\bibitem{Lai:2009ke}
K.~C.~Lai, G.~L.~Lin and T.~C.~Liu,
%``Determination of the Neutrino Flavor Ratio at the Astrophysical Source,''
Phys.\ Rev.\ D {\bf 80}, 103005 (2009).
%arXiv:0905.4003 [hep-ph].
%%CITATION = ARXIV:0905.4003;%%
%
\bibitem{Pakvasa:2004hu}
For a review on neutrino flavor ratios in various astrophysical
sources, see S.~Pakvasa,
%``Neutrino properties from high energy astrophysical neutrinos,''
Mod.\ Phys.\ Lett.\  A {\bf 19}, 1163 (2004) [Yad.\ Fiz.\  {\bf 67},
1179 (2004)].
%[hep-ph/0405179].
%%CITATION = YAFIA,67,1179;%%
%
\bibitem{GonzalezGarcia:2007ib}
M.~C.~Gonzalez-Garcia and M.~Maltoni,
%``Phenomenology with Massive Neutrinos,''
Phys.\ Rept.\  {\bf 460}, 1 (2008).
%[arXiv:0704.1800 [hep-ph]].
%%CITATION = PRPLC,460,1;%%
%
%
\bibitem{lhl_2010}
T.~C.~Liu, M.~A.~Huang and G.~.L.~Lin, arXiv:1004.5154 [hep-ph].
%
\bibitem{Balantekin:1999dx}
A.~B.~Balantekin and G.~M.~Fuller,
%``Constraints on neutrino mixing,''
Phys.\ Lett.\  B {\bf 471}, 195 (1999).
%[arXiv:hep-ph/9908465].
 %%CITATION = PHLTA,B471,195;%%
%
\bibitem{Harrison:2002et}
 P.~F.~Harrison and W.~G.~Scott,
 %``mu - tau reflection symmetry in lepton mixing and neutrino oscillations,''
 Phys.\ Lett.\  B {\bf 547}, 219 (2002).
 %[arXiv:hep-ph/0210197].
 %%CITATION = PHLTA,B547,219;%%
 %
\bibitem{Allison:2009rz}
P.~Allison {\it et al.},
%``IceRay: An IceCube-centered Radio-Cherenkov GZK Neutrino Detector,''
Nucl.\ Instrum.\ Meth.\  A {\bf 604}, S64 (2009).
% [arXiv:0904.1309 [astro-ph.HE]].
%%CITATION = NUIMA,A604,S64;%%
%
\bibitem{Lai:2010tj}
K.~C.~Lai, G.~L.~Lin and T.~C.~Liu,
%``Flavor Transition Mechanisms of Propagating Astrophysical Neutrinos -A
%Model Independent Parametrization,''
Phys.\ Rev.\  D {\bf 82}, 103003 (2010).
% [arXiv:1004.1583 [hep-ph]].
  %%CITATION = PHRVA,D82,103003;%%
\end{thebibliography}
\end{document}